\documentclass{elegantpaper}
\usepackage[utf8]{inputenc}

\title{Producing Histopathology Phantom Images using Generative Adversarial Networks to improve Tumor Detection}
\author{Vidit Gautam$^1$ \and $^1$Pioneer Academics, Emory University Department of Computer Science}
\date{Sept 11, 2020}

\begin{document}

\maketitle

\keywords{Generative Adversarial Networks,
            Histopathology,
            Medical Imaging,
            Deep Learning\\[2ex]}

\begin{abstract}
    Advance in medical imaging is an important part in deep learning research. One of the goals of computer vision is development of a holistic, comprehensive model which can identify tumors from histology slides obtained via biopsies. A major problem that stands in the way is lack of data for a few cancer-types.  In this paper, we ascertain that data augmentation using GANs can be a viable solution to reduce the unevenness in the distribution of different cancer types in our dataset. Our demonstration showed that a dataset augmented to a 50$\%$ increase causes an increase in tumor detection from 80$\%$ to 87.5$\%$.\\[2ex]
    
\end{abstract}

\section{Introduction}

Cancer treatment used to be based on ‘benign’ or ‘malignant’ before. But as oncology flourished and the term ‘cancer’ enveloped over 300 different tumor types, identification of the characteristic morphology for each tumor became critical. Cancer cannot be conclusively diagnosed without a biopsy. Only after a complete diagnosis done by a surgical pathologist can a doctor develop a plan for treatment. A pathologist’s analysis entails details of the type and origin of tumor, level of anaplasia, level of invasion, the numbers of lymph nodes with and without the tumor, enzyme activity, ploidy etc. This depends on the type of cancer. It requires a gross and introspective examination of structure, careful coordination among different specialisations (cytological pathology, clinical pathology, surgical pathology), and close collaboration with the caring clinician to synthesise information and create a pathology report. \cite{connolly2003role} \cite{leong2011changing} \par 

Due to an increasing workload in the field of cancer related diseases along with a decrease in the number of pathologists, automated assistance will be of great significance in the near future. \cite{petriceks2018trends} \cite{329a9967fc334d9caa48efc998bd6d16} This is where Deep Learning takes stage. We can support a pathologist’s daily routine by facilitating clinical practice with computer-based processing and image analysis. \par

Diagnosis from biopsy can be aided using Deep Learning models. For example, deep learning models can identify cancerous tissue against non-cancerous tissue, determine low level features such as percentage tumor area, percentage of cells in mitotic phases or the presence of hormone receptors. It is also possible for more sophisticated models to further identify high level features described previously such as anaplasia level, level of invasion or enzyme activity. \par
In some areas computation is more effective that current manual methods. For example, counting positively and negatively stained cells under Immunohistochemical Staining (IHC). To provide an IHC interpretation, pathologists provide estimates of positive/negative stained cells which suffer from poor reproducibility. This is something which can easily be automated. [4]  \par
Whole-slide scanners can digitise entire histology slides without much effort. These can generate vast amounts of digital data which opens up avenues for training Deep Learning models to fulfil analytical tasks in oncopathology. \cite{farahani2015whole} One such task is tumor detection, a process which, when automated, would greatly increase the efficiency of pathologists. Researchers have focused on the identification particular tumor types \cite{abdel2016breast}\cite{araujo2017classification}\cite{bejnordi2018using}\cite{coudray2018classification}, but a comprehensive, fully realised model would be capable of identifying any tumor type with high accuracy. This kind of model requires large diverse datasets for every cancer type. Immediately a problem rises as hospitals do not receive cancer patients with a uniform distribution of cancer types and thus cannot collect enough data for rare cancer types. \par
In this paper, we test the functionality of Generative Adversarial Networks (GAN) as a reasonable solution to augment data of rare cancers. Our goal is to test whether data augmentation using GANs can improve tumor detection models. \par

Due to an increasing workload in the field of cancer-related diseases along with a decrease in the number of pathologists, automated assistance will be of great significance in the near future. [18] [16] This is where Deep Learning takes the stage. We can support a pathologist’s daily routine by facilitating clinical practice with computer-based processing and image analysis.
Diagnosis from a biopsy can be aided using Deep Learning models. For example, deep learning models can identify cancerous tissue against non-cancerous tissue, determine low-level features such as percentage tumor area, percentage of cells in mitotic phases, or the presence of hormone receptors. It is also possible for more sophisticated models to further identify high-level features described previously such as anaplasia level, level of invasion, or enzyme activity.
In some areas, computation is more effective than the current manual methods. For example, counting positively and negatively stained cells under Immunohistochemical Staining (IHC). To provide an IHC interpretation, pathologists provide estimates of positive/negative stained cells that suffer from poor reproducibility. This is something that can easily be automated. [4]
Whole-slide scanners can digitize entire histology slides without much effort. These can generate vast amounts of digital data which opens up avenues for training Deep Learning models to fulfill analytical tasks in oncopathology. [9] One such task is tumor detection, a process which, when automated, would greatly increase the efficiency of pathologists. Researchers have focused on the identification of particular tumor types [1][2][3][7], but a comprehensive, fully realized model would be capable of identifying any tumor type with high accuracy. This kind of model requires large diverse datasets for every cancer type. Immediately a problem rises as hospitals do not receive cancer patients with a uniform distribution of cancer types and thus cannot collect enough data for rare cancer types.\par

\section{Background and Related Work}

A wide range of image augmentation techniques have been used in machine learning research, including flipping, rotating, shearing, cropping, zooming in/out, changing brightness, perturbing, texture transfer, style transfer , CNN based approaches, and GANs. \cite{mikolajczyk2018data} \par

The original GAN structure was developed by (Goodfellow et al.) \cite{goodfellow2014generative}. Since then, many alterations have been made to suit different purposes. During the development of PathologyGAN, (Shaban et al.)\cite{quiros2019pathology} included a mapping layer before the generator, which structures the unstructured latent space, controlling the features produced by the generator. (Huo et al.) \cite{huo2018adversarial} removes the generator entirely and adds an encoder-decoder structure in its stead, allowing for images to be translated from one domain to another. The application of this in Medical Imaging is Image Translation between MRIs, CT scans and PET scans. A similar effect can be gained by CycleGANs, which remove the discriminator and instead calculate Cycle Consistency Loss from two generators attempting to generate images from domain X1 to domain X2 and vice versa (Zhu et al.) \cite{xu2014weakly}. cGANs add classification into the structure of GANs, so generators can produce features based on classification and discriminators can classify the training dataset. \par

Since the genesis of GANs, researchers have had difficulty evaluating these models. Many different metrics such as Frechet Inception Distance (FID) \cite{heusel2017gans}, Maximum Mean Discrepancy (MMD)\cite{gretton2012kernel}, 1-Nearest Neighbor classifier (1-NN)  \cite{lopez2016revisiting}\cite{gretton2012kernel}, Kernel Inception Distance (KID) (Binkowski et al., 2018)\cite{binkowski2018demystifying}, and studies such as (Xu et al., 2018; Barratt and Sharma, 2018)’s work \cite{xu2014weakly} have described their advantages and disadvantages. \par

A wide range of image augmentation techniques have been used in machine learning research, including flipping, rotating, shearing, cropping, zooming in/out, changing brightness, perturbing, texture transfer, style transfer, CNN based approaches, and GANs. [17]
The original GAN structure was developed by (Goodfellow et al.). Since then, many alterations have been made to suit different purposes. During the development of PathologyGAN, (Shaban et al.)[19] included a mapping layer before the generator, which structures the unstructured latent space, controlling the features produced by the generator. (Huo et al.) [13] removes the generator entirely and adds an encoder-decoder structure in its stead, allowing for images to be translated from one domain to another. The application of this in Medical Imaging is Image Translation between MRIs, CT scans, and PET scans. A similar effect can be gained by CycleGANs, which remove the discriminator and instead calculate Cycle Consistency Loss from two generators attempting to generate images from domain X1 to domain X2 and vice versa (Zhu et al.) [22]. cGANs add classification into the structure of GANs, so generators can produce features based on classification and discriminators can classify the training dataset.
Since the genesis of GANs, researchers have had difficulty evaluating these models. Many different metrics such as Frechet Inception Distance (FID) [12], Maximum Mean Discrepancy (MMD)[10], the 1-Nearest Neighbor classifier (1-NN) [15][10], Kernel Inception Distance (KID) (Binkowski et al., 2018)[4], and studies such as (Xu et al., 2018; Barratt and Sharma, 2018)’s work [22] have described their advantages and disadvantages.\par

The attitude of machine learning researchers for digital pathology is to build classifiers which achieve pathologist-level diagnoses for a few cancer types. (Esteva et al., 2017; Wei et al., 2019; Han et al., 2017)\cite{estava2017dermatologist}\cite{han2017breast} The primary goal here is to aid decision by computer-human interaction. (Cai et al., 2019)\cite{cai2019human}
There has also recently been interest in utilising GANs for digitised staining (Rana et al., 2018; Xu et al., 2019)\cite{rana2018computational} (Ghazvinian Zanjani et al., 2018) \cite{zanjani2018stain}, phantom image generation (Senaras et al. 2018) \cite{senaras2018optimized} and nuclei segmentation. \par

The attitude of machine learning researchers for digital pathology is to build classifiers that achieve pathologist-level diagnoses for a few cancer types. (Esteva et al., 2017; Wei et al., 2019; Han et al., 2017)[8][11] The primary goal here is to aid decision by computer-human interaction. (Cai et al., 2019)[5] There has also recently been interest in utilizing GANs for digitized staining (Rana et al., 2018; Xu et al., 2019)[20] (Ghazvinian Zanjani et al., 2018), phantom image generation (Senaras et al. 2018) [21] and nuclei segmentation.\par

\section{Dataset}

To train our model, a H$\&$E Breast Cancer dataset was selected. The dataset is derived from 162 whole-slide images leading to 277,524 50x50 patches of images. Of these, we used 3324 non-tumor patches and 4289 tumorous patches to train the GAN and Convolutional Network. These patches were selected randomly from the whole dataset.  \par

\section{Proposed Architecture}

\begin{figure}[h]
    \centering
    \includegraphics[scale=0.4]{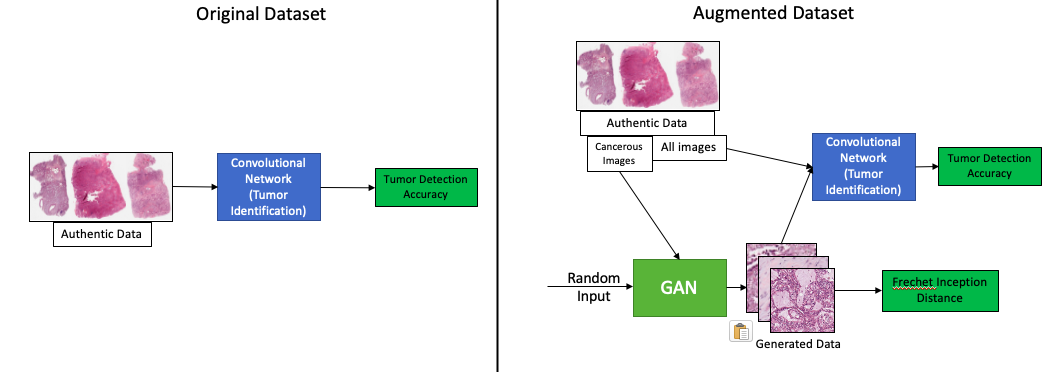}
    \caption{Overarching Structure: The Original Dataset \textit{$F_o$} is input into the Convolutional Network and the Augmented Dataset \textit{$F_a$} is input into the Convolutional Network}
    \label{Overarching Structure}
\end{figure}

To test the improvement in tumor detection with data augmentation, we obtain tumor-detection accuracy twice. Once, with the original dataset, and once augmented with generated images. You can see the overarching structure of networks in the Figure 1. \par

\subsection{Generative Adversarial Network}

Generative Adversarial Networks contain two submodels: the generator (G) and the discriminator (D). The generator model takes in a random vector, \textit{z} as input, and outputs fake histology images, \textit{$x_g$}. \textit{z} is purely random noise, based on the distribution \textit{p(z)} which, for simplicity, we chose as a uniform distribution. \textit{$x_g$} is going to be trained to be similar to real images \textit{$x_r$}, drawn from the real data \textit{$p_r(x)$}.  \par
The input to D is either \textit{$x_g$} or \textit{$x_r$}. The output of D, \textit{$y_1$} is a single value indicating whether the sample is ‘real’ or ‘fake’. Optionally, Discriminator loss and Generator loss are also output when D and G respectively are being trained. After successful training, generated samples form a distribution \textit{$p_g(x)$}, which is approximately the distribution of real images \textit{$p_r(x)$}. \par

The discriminator and generator trained using Adam Optimizer with $\beta_1 = 0.5$. While the discriminator’s goal is to identify which image is real, the generator goal is to confuse the discriminator by producing realistic images. \par
The discriminator’s loss function is output from itself as it is used in training to improve accuracy at identifying real images. However, the generator’s loss function is also output from the discriminator since its goal is to essentially fool the discriminator by producing realistic images. The generator’s training is complete once the discriminator cannot consistently identify fake images i.e. the probability that a generated image is classified as ‘fake’ is 50{$\%$}. Then, either the discriminator is trained to improve at identifying fake images or the GAN is considered to be fully trained. The training goals of D and G can be expressed as: \par

\begin{equation}
  \min_{G}\max_{D}{V\left(D,G\right)}=\mathbb{E}_{x~p_{data}}\left[\log{D\left(x\right)}\right]+\mathbb{E}_{z~p_z\left(z\right)}\left[\log{\left(1-D\left(G\left(z\right)\right)\right)}\right]
\end{equation}

Once the GAN is trained and it synthesises phantom images, the difference between the two distributions \textit{$p_g(x)$} and \textit{$p_r(x)$} is calculated as Frechet Inception Distance (FID). This characterises the effectiveness of the GAN in symbolising the original dataset when generating images. \par

\subsection{Convolutional Neural Network}

The Convolutional Network is utilised two times, once with the original dataset \textit{$F_o$} and once with the augmented dataset \textit{$F_a$}. Batches of 100 images are input into into the model, and then feature detection maps are utilised to output a classification of each image as Cancerous and Non-Cancerous. We use the Adagrad Optimizer to train this model.  \par

\section{Results}

\begin{table}[h]
    \centering
    \begin{tabular}{||c|c||}
        \hline
        Dataset & H\&E Breast Cancer IDC Tissue  \\
        Frechet Inception Distance & 35.495 \\
        Augmented Dataset Percentage Increase & 50.00$\%$ \\
        Original Convolutional Network Accuracy & 80.00$\%$ \\
        Augmented Convolutional Network Accuracy & 87.00$\%$ \\
        Convolutional Network Accuracy Percentage Increase & 08.75$\%$ \\
        \hline
    \end{tabular}
    \caption{The augmented dataset improves convolutional accuracy by 8.75$\%$}
    \label{result_table}
\end{table}

Results can be seen in Table 1. With a FID of 35.495, the GAN can produce a reliable distribution of synthesised images for the H\&E Invasive Ductal Carcinoma Dataset. Table 1 also represents the percentage increase percentage accuracy. As the dataset was augmented by 50$\%$, tumor detection accuracy increased by 8.75$\%$. Hence, we can state that GANs can cause a percentage increase accuracy of tumor detection models.

However, there are limitations in the proposed method. Out testing happened on a single dataset, which has been derived from a diverse selection of data. This is not representative of the type of data that would be obtained for rare cancer types, as subjects would be scarce. 

In future research, we anticipate that a wide variety of datasets would be tested with GANs for increase in tumour detection accuracy, or other models executing pathology analysis. We also expect different types of GANs such as BigGAN, StyleGAN or PathologyGAN to be used to produce images since these provide desired advantages such as high resolution or structured histology images. \par

\subsection{Conclusions}
This paper contemplates the problem of inconsistency in data collection for rare cancers and its effects on future deep learning research. The solution proposed is to use Generative Adversarial Networks, which can generate fake images from a small dataset. Our demonstration proves that GANs can be effective as a data augmentation strategy for Deep Learning research in Digital Pathology. Looking forward, GANs can serve as a crucial factor to equalize the distribution of different cancer-types when developing holistic Deep Learning Models. \par

\section*{Acknowledgements}
To Professor Avani Wildani, for assisting me throughout this research experience and teaching me about Biological Neural Networks.

\printbibliography[heading=bibintoc, title=\ebibname]

\end{document}